\documentclass[12pt]{article}
\usepackage{geometry} 
\usepackage[english]{babel}
\usepackage[english]{layout}

\usepackage{mathpazo} 
\usepackage[scaled]{helvet} 
\usepackage{courier} 
\normalfont
\usepackage[T1]{fontenc}

\usepackage{array} 
\usepackage{paralist} 
\usepackage{verbatim} 

\usepackage[utf8]{inputenc} 
\usepackage{textcomp} 
\usepackage{graphicx}  
\usepackage{epstopdf} 
\usepackage{flafter}  

\usepackage{amsfonts,latexsym,amsthm,amssymb}
\usepackage{amsmath,amssymb,bbm}  
\usepackage{bm}  

\usepackage{memhfixc}  
\usepackage[activate]{pdfcprot}  
\usepackage{fancyhdr} 
\usepackage{sectsty}
\allsectionsfont{\sffamily\mdseries\upshape} 

\numberwithin{equation}{section}

\def\Ref{\ref}
\def\c{\cite}
\def\Ref#1{(\ref{#1})}

\title{Symmetries of spin systems and Birman-Wenzl-Murakami algebra}

\author{ \textsf{P. ~P.~Kulish,}
\thanks{E-mail address: kulish@euclid.pdmi.ras.ru}
\textsf{ ~~N. ~Manojlovi\'c}
\thanks{E-mail address: nmanoj@ualg.pt}
\textsf{ and Z. ~Nagy}
\thanks{E-mail address: 
zoltan.nagy@m4x.org} \\
\\
\textit{$^{\ast}$St. Petersburg Department of Steklov Mathematical Institute} \\
\textit{Fontanka 27, 191023, St. Petersburg, Russia} \\
\\
\textit{$^{\ast \dag\ddag}$
Grupo de F\'{\i}sica Matem\'atica da Universidade de Lisboa} \\
\textit{Av. Prof. Gama Pinto 2, PT-1649-003 Lisboa, Portugal} \\
\\
\textit{$^{\dag}$Departamento de Matem\'atica, F. C. T.,
Universidade do Algarve}\\ 
\textit{Campus de Gambelas, PT-8005-139 Faro, Portugal}}
\date{}

\begin{document}

\maketitle
\thispagestyle{empty}

\begin{abstract}
We consider integrable open spin chains related  to the quantum affine algebras $\mathcal{U}_q(\widehat{o(3)})$ and $\mathcal{U}_q(A_2^{(2)})$. We discuss the symmetry algebras of these chains with the local $\mathbb{C}^3$ space related to the Birman-Wenzl-Murakami algebra. The symmetry algebra and the Birman-Wenzl-Murakami algebra centralize each other in the representation space $\mathcal{H} = \otimes _{1}^N \mathbb{C}^3$ of the system, and this determines the structure of the spin system spectra. Consequently, the corresponding multiplet structure of the energy spectra is obtained.
\end{abstract}

\clearpage 
\newpage

%
%
\section{Introduction}

The development of the quantum inverse scattering method (QISM) \c{TakhFadI, Faddeev, KS} as an approach to construction and exact solution of quantum integrable systems has lead to the foundations of the theory of quantum groups \c{Drinfeld,Jimbo}. The representation theory of quantum groups is naturally connected to the spectral theory of the integrals of motion of quantum systems. In particular, this connection appeared in the combinatorial approach to the question of completeness of the eigenvectors of the $XXX$ Heisenberg spin chain \c{TakhFadII}. 

Important properties of quantum integrable systems are related with their symmetry algebra and are defined by a bigger algebra which gives the main relations underlining integrability, the so-called RLL-relations \c{TakhFadI}. In the case of most known isotropic Heisenberg chain of spin $1/2$ (XXX-model) the symmetry algebra is $sl_2$, the Hamiltonian is an element of the group algebra $\mathbb{C}[\mathfrak{S}_N]$ of the symmetric group $\mathfrak{S}_N$. The fundamental relations of the auxiliary L-matrix entries generate an infinite dimensional quantum algebra -- the Yangian $\mathcal{Y}(sl_2)$. The actions of $sl_2$ and  $\mathfrak{S}_N$ on the state of space $\mathcal{H} = \otimes _{1}^N \mathbb{C} ^2$ are mutually commuting (the Schur-Weyl duality). Extension of this scheme to a particular case of the Hecke algebra -- the Temperley-Lieb algebra, instead of the symmetric group and corresponding new quantum algebras were proposed in \c{Kulish, KMN}. Here we consider a further generalization -- the case of the Birman-Wenzl-Murakami algebra \c{BW} and its specific representations in $\mathbb{C}^3\otimes \mathbb{C}^3$ given by the spectral parameter dependent R-matrices. These R-matrices correspond to different quantum affine algebras $\mathcal{U}_q(\widehat{o(3)})$, $\mathcal{U}_q(A^{(2)}_2)$, $\mathcal{U}_q(\widehat{osp(1|2)})$ and $\mathcal{U}_q(sl(1|2)^{(2)})$. Although corresponding spin systems were analysed in a variety of papers (detailed references are given below) we point out the connection of the open spin chains with the Birman-Wenzl-Murakami algebra as a centralizer of the symmetry algebra. 

For the XXZ-model of spin 1 the appropriate dynamical symmetry algebra is $\mathcal{U}_q(\widehat{o(3)})$ and its symmetry algebra is $\mathcal{U}_q(o(3))$ \cite{Nepomechie}. The corresponding R-matrix was found in \c{ABZVAF}, see also \c{Jimbo}, and it can also be obtained by the fusion procedure starting from the R-matrix of the XXZ-model of spin \textonehalf \ \cite{KS}. 

The R-matrix of $\mathcal{U}_q(A^{(2)}_2)$ in $\mathbb{C}^3 \otimes \mathbb{C}^3$ was found in \c{IzerginKorepin} and the corresponding periodic spin chain was solved by recurrence algebraic Bethe ansatz in \c{Tarasov}. 

These two spectral parameter dependent R-matrices are the two versions of the Yang-Baxterization procedure for a given representation of the Birman-Wenzl-Murakami algebra $W_2(q, \nu = q^{-2})$ in $\mathbb{C}^3 \otimes \mathbb{C}^3$\cite{Jones,Ge,Isaev}.

The two additional R-matrices related to the quantum affine super-algebras can be obtained by considering the Birman-Wenzl-Murakami algebra $W_2(-q, \nu = -q^{-2})$ and taking into account the connection between the solutions of the Yang-Baxter equation and the solutions of the ($\mathbb{Z}_2$ graded) super-Yang-Baxter equation \c{KS1}. In this case the representation of the Birman-Wenzl-Murakami (BMW) algebra is the centralizer of the $\mathcal{U}_q(osp(1|2))$ action in the tensor product of its fundamental representation.  

We point out the multiplet structure of the energy spectra of the corresponding open spin chain Hamiltonians. The quantum determinant of the algebra $\mathcal{U}_q(A^{(2)}_2)$ is also given.

The symmetry properties of integrable spin chains depend also on the boundary conditions, for example, there are soliton preserving versus non-soliton preserving boundary conditions, see \cite{AACDFR} and the references therein. For the XXZ-chain of spin \textonehalf \ particular boundary conditions yield the spectrum of the system which has clear multiplet structure of the irreducible representations of the Hecke algebra $\mathcal{H}_N(q)$ and the symmetry algebra $\mathcal{U}_q(sl_2)$. However, there are also K-matrices defining the integrable boundary conditions of the XXZ model such that the whole space of states is just an irreducible representation of the reflection equation algebra.

The paper is organised as follows. In Section II the R-matrix of the model $XXZ_1$ and its properties are reviewed. The emphasis is given to its connection to the $\mathcal{U}_q(o(3))$ constant R-matrix and the corresponding realisation of the BMW algebra. In Section III the Izergin-Korepin R-matrix is reviewed along the same lines. It was shown that although the constant R-matrix is the same as in the case of the XXZ spin-1 however, the corresponding Yang -Baxterization of the BMW algebra generators yields different spectral parameter dependent R-matrix. In Section IV the definition of the  Birman-Wenzl-Murakami algebra is reviewed in general. Also, some properties of the symmetrizers and antisymmetrizers in the particular case of the BMW algebra $W_N(q, q^{-2} )$, corresponding to the $XXZ_1$ and $A^{(2)}_2$ R-matricies, are studied. The symmetries of the corresponding open spin chains are discussed in Section V. In particular, the realization of the BMW algebra as the centralizer of the symmetry algebra of the open spin chain is analysed. The multiplet structure of the energy spectra of the corresponding open spin chain Hamiltonians is the main result of this analysis.   
Our conclusions and directions for further research are given in the last Section. 


%
%
\section{R-matrix of XXZ spin-1chain}

Following \c{ABZVAF,KS1, KS, Jimbo}, the $9\times 9$ R-matrix of the XXZ-chain of spin one can be expressed as follows
\begin{equation}
\label{RXXZ}
R (\lambda,\eta) = 
\left( \begin{array}{ccc|ccc|ccc}
a_1& & & & & & & & \\
&a_2& &b_1 & & & & & \\
&  &a_3& &b_2& &b_3& & \\\hline
& c_1& &a_2& & & & & \\
& &c_2 & &a_4& &b_2& & \\
& & & & &a_2& &b_1 & \\ \hline
& &c_3& &c_2& &a_3& & \\
& & & & & c_1& &a_2& \\
& & & & & & & &a_1
\end{array} 
\right) ,
\end{equation}
where the functions are 
\begin{equation}
\notag
\left.\begin{array}{ll}
a_1= \sinh (\lambda +\eta) \sinh (\lambda +2\eta), & b_2 = e^{\lambda} \sinh \lambda \sinh 2\eta ,\\
a_2= \sinh \lambda  \sinh (\lambda +\eta) , & b_3 = e^{2\lambda} \sinh \eta \sinh 2\eta, \\
a_3= \sinh \lambda  \sinh (\lambda -\eta) , & c_1= e^{-\lambda} \sinh (\lambda +\eta) \sinh 2\eta,\\
a_4= \sinh \lambda  \sinh (\lambda +\eta) + \sinh \eta \sinh 2\eta, & c_2= e^{-\lambda} \sinh \lambda \sinh 2\eta ,\\
b_1= e^{\lambda} \sinh (\lambda +\eta) \sinh 2\eta, & c_3= e^{-2\lambda} \sinh \eta \sinh 2\eta .
\end{array}\right.
\end{equation}
The R-matrix satisfies the Yang-Baxter equation in the space $\mathbb{C}^3 \otimes \mathbb{C}^3 \otimes \mathbb{C}^3$
\begin{equation}
\label{YBE}
R_{12} ( \lambda ) R_{13} ( \lambda+ \mu) R_{23} ( \mu) = R_{23} ( \mu ) R_{13} ( \lambda + \mu) R_{23} ( \lambda),
\end{equation}
where we use the standard notation of the QISM \cite{TakhFadI, Faddeev, KS}.

This form of the  R-matrix is related with the symmetric one $R_{12}^t ( \lambda,\eta) = R_{12} ( \lambda, \eta)$ by the similarity transformation
\begin{equation}
R_{12} ( \lambda, \eta) \to \mathrm{Ad} \exp (\alpha\lambda (h_1-h_2)) R_{12} ( \lambda, \eta) ,
\end{equation}
with $\alpha =$ \textonehalf \ and $h = \mathrm{diag}(1,0,-1)$. The transformed R-matrix still obeys the Yang-Baxter equation due to the $U(1)$ symmetry of the initial R-matrix
\begin{equation}
\label{0-weght}
[h_1+h_2, R_{12} ( \lambda, \eta) ] = 0.
\end{equation}

The R-matrix \Ref{RXXZ} has a few important properties: regularity, unitarity,  PT-symmetry and crossing symmetry.
The regularity condition at $\lambda = 0$ reads
\begin{equation}
\label{regRXXZ}
R(0, \eta) =  \sinh(\eta) \sinh(2\eta) \mathcal{P},
\end{equation}
where $\mathcal{P}$ is the permutation matrix of $\mathbb{C}^3 \otimes \mathbb{C}^3$.
The unitarity relation is
\begin{equation}
\label{unRXXZ}
R_{12}( \lambda) R_{21}(-\lambda) = \rho ( \lambda) \mathbbm{1},
\end{equation}
here $R_{21}(\lambda) = \mathcal{P} R_{12}( \lambda) \mathcal{P}$ and $\rho$ is the following function
\begin{equation}
\label{normXXZ}
 \rho ( \lambda) = \sinh(\lambda+\eta) \sinh(\lambda+2\eta)  \sinh(\lambda-\eta)\sinh(\lambda-2\eta) .
 \end{equation}
The so-called PT-symmetry states
\begin{equation}
\label{ptRXXZ}
R _{12}^t (\lambda) =  R _{21}(\lambda).
\end{equation}
Finally, it has the following crossing symmetry property 
\begin{equation}
\label{R-crossingXXZ}
R ( \lambda) =  \left(Q \otimes \mathbbm{1}\right) R ^{t_2}( -\lambda -\eta ) \left(Q \otimes \mathbbm{1}\right) ,
\end{equation}
where $t_2$ denotes the transpose in the second space and the matrix $Q$ is given by
\begin{equation}
\label{Q-mat}
Q = \left(\begin{array}{ccc}0 & 0 & -e^{-\eta} \\0 & 1 & 0 \\- e^{\eta} & 0 & 0\end{array}\right).
\end{equation}

The R-matrix \Ref{RXXZ} in the braid group form 
\begin{equation}
\label{checkR}
\check{R} (\lambda,\eta) = \mathcal{P} R (\lambda,\eta) ,
\end{equation}
admits the spectral decomposition
\begin{align}
\label{chRXXZproj}
\check{R} (\lambda,\eta) 
&= \sinh (\lambda +\eta) \sinh (\lambda + 2\eta) P_5 (\eta) 
- \sinh (\lambda +\eta) \sinh (\lambda - 2\eta) P_3 (\eta) \notag \\
&+ \sinh (\lambda -\eta) \sinh (\lambda - 2\eta) P_1 (\eta) ,
\end{align}
here 
\begin{equation}
\label{P5}
P_5 (\eta) = \mathbbm{1} - P_3 (\eta) - P_1 (\eta),
\end{equation}
\begin{equation}
\label{P3}
\!P _3  (\eta) = \frac{1} {e^{2\eta} + e^{-2\eta}}
\left( \begin{array}{ccc|ccc|ccc}
0& & & & & & & & \\
&e^{2\eta}& &-1& & & & & \\
& &1& &\omega & &-1& & \\\hline
&-1&  &e^{-2\eta}&   & &  & & \\
& &\omega & &\omega ^2& &-\omega & & \\
& & & & &e^{2\eta}& &-1& \\\hline
& &-1& &-\omega & &1& & \\
& &  & &   &-1&  &e^{-2\eta}& \\
& &  & &   & &  & & 0
\end{array}  \right) ,
\end{equation}
here $\omega (e^{\eta}) = e^{\eta}-e^{-\eta}$ and 
\begin{equation}
\label{P1}
\!P _1 (\eta) = \frac {1} {e^{2\eta} + 1 + e^{-2\eta}}
\left( \begin{array}{ccc|ccc|ccc}
0& & & & & & & & \\
&0 & & & & & & & \\
&  &e^{2\eta}& &-e^{\eta}& &1& & \\\hline
& & &0 & & & & & \\
& &-e^{\eta}& &1& &-e^{-\eta}& & \\
& & & & &0 & & & \\ \hline
& &1& &-e^{-\eta}& &e^{-2\eta}& & \\
& & & & & & &0& \\
& & & & & & & &0
\end{array} 
\right) .
\end{equation}
These are projectors to the five, three and one dimensional eigenspace, respectively. Thus the R-matrix \Ref{RXXZ} has four degeneration points $\lambda = \pm \eta$, and  $\lambda = \pm 2\eta$. Its rank at $\lambda = \eta$ is eight, at $\lambda = 2\eta$ is five, at  $\lambda =  -2\eta$ is four and finally at  $\lambda = -\eta$ is one.

The R-matrix \Ref{checkR} can also be expressed in the following form, useful for the asymptotics   
\begin{equation}
\label{chRXXZasymp}
\check{R} (\lambda,\eta)  = \frac{e^{\eta}}{4} \left( e^{2\lambda} -1 \right) \check{R} (\eta) + \left( \sinh \eta \sinh 2\eta \right) \mathbbm{1} + \frac{e^{-\eta}}{4} \left( e^{-2\lambda} -1 \right)\check{R} ^{-1}(\eta).  
\end{equation}
A relevant observation is that the constant R-matrix  
\begin{equation}
\label{bgR}
\check{R} ^{\pm 1} (\eta) = \lim _{\lambda \to \pm \infty} \left( 4 \exp (\mp ( 2\lambda + \eta)) \check{R} (\lambda,\eta) \right)
\end{equation}
being a solution of the Yang-Baxter equation in the braid group form 
\begin{equation}
\label{BGYBe}
\check{R}_{12}\check{R}_{23}\check{R}_{12} = \check{R}_{23}\check{R}_{12}\check{R}_{23},
\end{equation}
has the spectral decomposition $(q=e ^{2\eta})$
\begin{equation}
\label{bgRspec}
\check{R} (\eta) = q P_5 (\eta) - \frac{1}{q} P_3 (\eta) + \frac{1}{q^2} P_1 (\eta) .
\end{equation}
Hence, $\check{R} (\eta)$ satisfies the cubic equation
\begin{equation}
\label{chRcub}
\left(\check{R} (\eta)  - q \mathbbm{1} \right) \left(\check{R} (\eta) + \frac{1}{q} \mathbbm{1} \right) \left(\check{R} (\eta) - \frac{1}{q^2} \mathbbm{1} \right)=0 .
\end{equation}
Consequently, its minimal polynomial is 
\begin{equation}
\label{chRpol}
(\alpha - q) (\alpha + \frac{1}{q})(\alpha - \frac{1}{q^2}).
\end{equation}
Its matrix form is 
\begin{equation}
\label{bgRmat}
\check{R} (\eta) = 
\left( \begin{array}{ccc|ccc|ccc}
e^{2\eta}& & & & & & & & \\
&0& &1& & & & & \\
& &0& & & &e^{-2\eta}& & \\\hline
&1&  &\omega&   & &  & & \\
& & & &1& &e^{-\eta}\omega & & \\
& & & & &0& &1& \\\hline
& &e^{-2\eta}& &e^{-\eta} \omega & &(1-e^{-2\eta}) \omega & & \\
& &  & & &1&  &\omega & \\
& &  & & & &  & & e^{2\eta}
\end{array}  \right),
\end{equation}
here $\omega (e^{2\eta}) = e^{2\eta}-e^{-2\eta}$.

For the purpose of establishing a relation with the Birman-Wenzl-Murakami algebra, the one dimensional projector $P_1 (\eta)$ is related to the rank one matrix $\mathcal{E} (\eta)$  
\begin{equation}
\label{matK}
\mathcal{E} (\eta) = \mu P_1 (\eta),
\end{equation}
with $\mu = q + 1 + 1/q$ and $q=e ^{2\eta}$. The matrix $\mathcal{E}(\eta)$ satisfies 
\begin{align}
\label{}
   \mathcal{E}^2 (\eta) &= \mu \mathcal{E} (\eta), \\
    \check{R}(\eta) \mathcal{E} (\eta)&=   \mathcal{E} (\eta)\check{R}(\eta) = \frac{1}{q^2} \mathcal{E}(\eta),
\end{align}
and also
\begin{equation}
\label{bmw}
\check{R} (\eta) - \check{R}^{-1} (\eta) = \omega (q) \left( \mathbbm{1}  - \mathcal{E} (\eta) \right) , 
\end{equation}
where $\omega (q) = q - 1/q$. From these relations we conclude that $\check{R}, \check{R}^{-1}$ and $\mathcal{E}$ provide a realisation of the Birman-Wenzl-Murakami algebra $W_N(q, 1/q^2 )$ \c{BW} in the space $\mathcal{H} = \otimes _{1}^N \mathbb{C}^3$. 

The projector $P_5(\eta)$ on five dimentional subspace of 
$\mathbb{C}^3\otimes\mathbb{C}^3$ corresponds to a symmetrizer of spin 2 
irreducible representation of the quantum algebra $\mathcal{U}_q(o(3))$. 
It can be used to construct an R-matrix for higher spin
$R^{(2,1)}(\lambda,\eta)\in\text{End}(\mathbb{C}^5\otimes\mathbb{C}^3)$ 
by the fusion procedure [3]
\begin{equation}
R^{(2,1)}(\lambda,\eta)\simeq\check{R}_{12}(2\eta,\eta)R_{13}(\lambda+\eta,\eta)
R_{23}(\lambda-\eta,\eta).
\end{equation}
It will be shown in Sec. V that one can use higher symmetrizers of the BMW-algebra $W_s(q,1/q^2)$ to get R-matrices
$R^{(s,1)}(\lambda,\eta)\in\text{End}(\mathbb{C}^{(2s+1)}\otimes\mathbb{C}^3)$, in this notation the original R-matrix is  $R^{(1,1)}(\lambda,\eta)$.

%
%
\section{Izergin-Korepin R-matrix}

Following \c{Tarasov}, the Izergin-Korepin R-matrix is expressed as follows
\begin{equation}
\label{RA22}
R (\lambda,\eta) = 
\left( \begin{array}{ccc|ccc|ccc}
a_1& & & & & & & & \\
&a_2& &b_1 & & & & & \\
&  &a_3& &b_2& &b_3& & \\\hline
& c_1& &a_2& & & & & \\
& &c_2 & &a_4& &b_2& & \\
& & & & &a_2& &b_1 & \\ \hline
& &c_3& &c_2& &a_3& & \\
& & & & & c_1& &a_2& \\
& & & & & & & &a_1
\end{array} 
\right) ,
\end{equation}
where the functions are 
\begin{equation}
\notag
\!\!\!\!\!\!\!\!\!\!\!\!\!\!\!\!\!\!
\left.\begin{array}{ll}
a_1= \sinh (\lambda-5\eta) +  \sinh (\eta), & b_2 = e^{2\eta} \sinh2 \eta \left( 1-e^{-\lambda}\right),\\
a_2= \sinh (\lambda-3\eta) +  \sinh (3\eta), & b_3 = -2e^{-\lambda+2\eta} \sinh \eta \sinh2 \eta - e^{-\eta} \sinh 4\eta , \\
a_3= \sinh (\lambda-\eta) +  \sinh (\eta), & c_1= -\sinh (2\eta)\left( e^{\lambda-3\eta} + e^{3\eta}\right),\\
a_4= \sinh (\lambda-3\eta) +  \sinh (3\eta)-\sinh (5\eta) +  \sinh (\eta), & c_2=  e^{-2\eta} \sinh2 \eta\left( 1-e^{\lambda}\right),\\
b_1= -\sinh (2\eta)\left( e^{-\lambda+3\eta} + e^{-3\eta}\right), & c_3= 2e^{\lambda-2\eta} \sinh \eta \sinh2 \eta - e^{\eta} \sinh 4\eta .
\end{array}\right.
\end{equation}
Like in the case of XXZ spin 1, this R-matrix \Ref{RA22} has four important properties: regularity, unitarity,  PT-symmetry and crossing symmetry.
The regularity condition at $\lambda = 0$ reads
\begin{equation}
\label{regRA22}
R(0, \eta) =  (\sinh(\eta) -\sinh(5\eta)) \mathcal{P},
\end{equation}
where $\mathcal{P}$ is the permutation matrix of $\mathbb{C}^3 \otimes \mathbb{C}^3$. The unitarity relation is
\begin{equation}
\label{unRA22}
R_{12}( \lambda) R_{21}(-\lambda) = \rho ( \lambda) \mathbbm{1},
\end{equation}
and $\rho$ is the following function
\begin{equation}
\label{normA22}
 \rho ( \lambda) = - (\sinh(\lambda+5\eta) -\sinh(\eta) )(\sinh(\lambda-5\eta) + \sinh(\eta)).
 \end{equation}
The so-called PT-symmetry states that the transpose of the R-matrix \Ref{RA22} is equal to the same R-matrix conjugated by the permutation matrix $\mathcal{P}$, that is
\begin{equation}
\label{ptRA22}
R _{12}^t (\lambda) =  R _{21}(\lambda).
\end{equation}
Also, the R-matrix \Ref{RA22} has the following crossing symmetry \c{Reshetikhin}
\begin{equation}
\label{R-crossingA22}
R ( \lambda) =  \left(Q \otimes \mathbbm{1}\right) R^{t_2} ( -\lambda +6\eta +  \imath \pi ) \left(Q \otimes \mathbbm{1}\right),
\end{equation}
where $t_2$ denotes the transpose in the second space and the matrix $Q$ is given in \Ref{Q-mat}.

In the braid group form the R-matrix \Ref{RA22}
\begin{equation}
\label{checkRA22}
\check{R} (\lambda,\eta) = \mathcal{P} R (\lambda,\eta) ,
\end{equation}
admits the spectral decomposition
\begin{align}
\label{chRA22proj}
\check{R} (\lambda,\eta) 
&= (\sinh (\lambda - 5\eta) + \sinh (\eta)) P_5 (\eta) 
- (\sinh (\lambda -\eta) + \sinh (5\eta)) P_3 (\eta) \notag \\
&+ (\sinh (\lambda +\eta) - \sinh (5\eta)) P_1 (\eta) ,
\end{align}
where the projectors $P_5 (\eta)$ , $P_3 (\eta)$ and $P_1 (\eta)$ are the same as in the equation \Ref{chRXXZproj} and are given in (\ref{P5}-15), respectively. Thus the R-matrix \Ref{RA22} has four degeneration points \c{Reshetikhin} $\lambda = \pm 4 \eta$, and  $\lambda = \pm (6\eta + \imath \pi )$. Its rank at $\lambda = - (6\eta + \imath \pi )$ is eight, at $\lambda = - 4\eta$ is six, at  $\lambda =  4\eta$ is three and finally at  $\lambda = 6\eta + \imath \pi$ is one.

The R-matrix \Ref{checkRA22} can also be expressed in the following form
\begin{align}
\label{chRA22asymp}
\check{R} (\lambda,\eta)  &= \frac{e^{3\eta}}{2} \left( 1- e^{-\lambda} \right) \check{R} (\eta) - \frac{1}{2}\left( e^{3\eta} + e^{-3\eta} \right) \left( e^{2\eta} -e^{-2\eta} \right)  \mathbbm{1} \notag \\
&- \frac{e^{-3\eta}}{2} \left( 1-e^{\lambda} \right)\check{R} ^{-1}(\eta),  
\end{align}
where the constant R-matrix used here is given in \Ref{bgRspec} and is the same as the one used in \Ref{chRXXZasymp}. This constant R-matrix, as it was pointed out, defines a representation of the BMW algebra $W_N(q, 1/q^2)$ in 
$\mathcal{H} = \otimes_1^N \mathbb{C}^3$. To confirm this, in the next section, we briefly review basic facts of the Birman-Wenzl-Murakami algebra.

The matrix $\check{R} (\lambda, \eta)$ \Ref{chRA22proj} at degeneration point $\lambda = 4 \eta$ is proportional to the rank 3 projector $P_3(\eta)$ \Ref{P3} which is a q-analogue of the antisymmetrizer on $\mathbb{C}^3 \otimes \mathbb{C}^3$. One can  further obtain the antisymmetrizer on $\mathbb{C}^3 \otimes \mathbb{C}^3 \otimes \mathbb{C}^3$ according to the fusion procedure [3]
\begin{equation}
\label{q-detA3}
\mathcal{A}_3 \simeq \check{R}_{12} (4\eta, \eta) \check{R}_{23} (8\eta, \eta) \check{R}_{12}(4\eta, \eta) .
\end{equation}
This matrix $\mathcal{A}_3 \in \mathrm{End}((\mathbb{C}^3)^{\otimes 3})$ has rank one. It can also be used to define a quantum determinant $\text{q-det} L(\lambda)$ of operator valued L-matrix $L(\lambda)$ satisfying the so-called RLL-relation, a milestone of the QISM,
\begin{equation}
\label{RLLlambda}
\check{R}_{12}(\lambda - \mu) L_1(\lambda)L_2(\mu) = L_1(\mu)L_2(\lambda) \check{R}_{12}(\lambda - \mu).
\end{equation}
In this case, the quantum determinant 
\begin{equation}
\label{q-detL}
\text{q-det} L(\lambda) \simeq \check{R}_{12} (4\eta, \eta) \check{R}_{23} (8\eta, \eta) \check{R}_{12}(4\eta, \eta) L_1(\lambda)L_2(\lambda - 4\eta) L_3(\lambda - 8\eta)
\end{equation}
is given by 
\begin{align}
\label{q-detL1}
&\text{q-det} L(\lambda) =   A_1(\lambda) C_1(\lambda -4\eta) C_3 (\lambda -8\eta)
-e^{-2\eta}A_1(\lambda) C_3(\lambda -4\eta) C_1(\lambda -8\eta)\notag \\
&-e^{-2\eta}C_1(\lambda) A_1(\lambda -4\eta) C_3(\lambda -8\eta) -e^{-2\eta} \omega(e^{\eta})C_1(\lambda) C_1(\lambda -4\eta) C_1(\lambda -8\eta)
\notag \\
&+e^{-2\eta}C_1(\lambda) C_3(\lambda -4\eta) A_1(\lambda -8\eta) 
+e^{-2\eta}C_3(\lambda) A_1(\lambda -4\eta) C_1(\lambda -8\eta) \notag \\
&-e^{-4\eta}C_3(\lambda) C_1(\lambda -4\eta) A_1(\lambda -8\eta) ,
\end{align}
where $A_i(\lambda), B_i(\lambda)$ and $C_i(\lambda), i = 1, 2, 3,$ are the operator entries of the L-matrix
\begin{equation}
L(\lambda) = \left(\begin{array}{ccc}A_1 & B_1 & B_3 \\C_1 & A_2 & B_2 \\C_3 & C_2 & A_3\end{array}\right).
\end{equation}
The vector in $(\mathbb{C}^3)^{\otimes 3}$ defining the rank one antisymmetrizer \Ref{q-detA3}, coinsides with the quantum completely antisymmetric tensor of \cite{Fiore}. It can be shown that the quantum determinant \Ref{q-detL} is central, with respect to the RLL-relation \Ref{RLLlambda}, due to the proportionality of the R-marix quantum determinant to the identity matrix
\begin{equation}
\text{q-det}_0 R_{01}(\lambda,\eta) \simeq \mathbbm{1} _1 \in  \mathrm{End}(\mathbb{C}^3).
\end{equation}
The $\text{q-det} L(\lambda)$ has a group-like property
\begin{equation}
\text{q-det} \left(L_{02}(\lambda) L_{01}(\lambda)\right) = \text{q-det} L_{02}(\lambda) \cdot \text{q-det}L_{01}(\lambda). 
\end{equation}

As the final remark in the discussion of the properties of the $XXZ_1$ and $A_2^{(2)}$ trigonometric R-matrices we point out that these R-matrices have different scaling limits. The $A_2^{(2)}$ R-matrix in the limit $\lambda \to \epsilon \lambda$, $\eta \to \epsilon \eta$ and $\epsilon \to 0$ yields the $sl(3)$-Yang R-matrix  
\begin{equation}
R (\lambda,\eta) = \lambda \mathbbm{1} - 4 \eta \mathcal{P},
\end{equation}
while in the $XXZ_1$ case the limit yields
\begin{equation}
R (\lambda,\eta) = \lambda (\lambda + \eta) \mathbbm{1} + 2\eta (\lambda + \eta) \mathcal{P} + 2 \lambda \eta \mathcal{K} ,
\end{equation}
where $\mathcal{K}$ is a rank 1 matrix, invariant with respect to the $O(3)$ transformations. 

In the quasi-classical limit $\eta \to 0$ these two trigonometric R-matrices also yield different classical r-matrices.

%
%
\section{Birman-Wenzl-Murakami algebra $W_N(q, \nu)$}

The defining  relations of the BMW algebra $W_N(q, \nu )$, for the generators $1$, $\sigma _i$ , $\sigma ^{-1}_i$ and $e_i$, $i = 1, \dots, N-1$, are recalled for convenience, \c{BW}
\begin{align}
\label{defBMW1}
\sigma _i \sigma _{i+1} \sigma _i  &=   \sigma _{i+1} \sigma _i  \sigma _{i+1} , \quad \sigma _i \sigma _j = \sigma _j \sigma _i, \ \mathrm{for} \ |i-j| > 1,\\
\label{defBMW2}
e_i  \sigma _i  &=   \sigma _i e_i = \nu e_i , \\
\label{defBMW3}
e_i  \sigma _{i-1}^{\pm 1}  e_i &= \nu ^{\mp 1} e_i  , \\
\label{defBMW4}
\sigma _i - \sigma ^{-1}_i &= \omega (q) (1-e_i),
\end{align}
where $\omega (q) = q - 1/q$. It can be shown that the dimension of the Birman-Wenzl-Murakami algebra $W_N(q, \nu)$ is $
 (2N-1)!!$ \cite{BW}. 
 
Many useful relations follow from the definition above, for example \c{Isaev}
\begin{equation}
\label{e-square}
{e_i}^2 = \mu e_i, \quad \mathrm{with} \quad \mu = \frac{\omega - \nu + 1/\nu}{\omega} = \frac{(q-\nu)(\nu + 1/q)}{\nu \omega}.
\end{equation}
Another important consequence of the relations (\ref{defBMW2},4) is a cubic relatoin for $\sigma _i$
\begin{equation}
(\sigma _i - q )  (\sigma _i + q^{-1} ) (\sigma _i - \nu ) = 0.
\end{equation}
There is the natural inclusion of $W_M(q, \nu) \subset W_N(q, \nu)$, $M<N$. Namely, the first $3(M-1)$ generators $\{ \sigma_i ^{\pm1}, e_i ;\  i = 1,2 \dots, M -1 \}$ of $W_N(q, \nu)$ define the algebra $W_M(q, \nu)$. 

The Yang-Baxterization procedure yields two spectral parameter dependent elements \c{Jones,Ge,Isaev}
\begin{equation}
\label{sigma(u)}
\sigma _i ^{(\pm)}(u) = \frac{1}{\omega} \left(u^{-1} \sigma _i - u \ \sigma _i ^{-1}\right) + \frac{\nu \pm q^{\pm 1}}{u\nu \pm q^{\pm 1} u^{-1}} e_i.
\end{equation} 
These elements satisfy the Yang-Baxter equation in the braid group form 
\begin{equation}
\label{bgYBe}
\sigma _i ^{(\pm)}(u) \sigma _{i+1} ^{(\pm)}(uv) \sigma _i ^{(\pm)}(v) =   \sigma _{i+1}^{(\pm)}(v)  \sigma _i ^{(\pm)}(uv)  \sigma _{i+1}^{(\pm)}(u). 
\end{equation}
Their unitarity relation is
\begin{equation}
\label{unitarity}
\sigma _i ^{(\pm)}(u) \sigma _i ^{(\pm)}(u^{-1}) = \left( 1 - \omega ^{-2} (u-u^{-1}) ^2\right). 
\end{equation}  
The regularity property of the Yang-Baxterized elements \Ref{sigma(u)} is important for the locality of Hamiltonian density of the corresponding spin chains and is valid on the algebraic level due to \Ref{defBMW4}. Also, these elements are normalised so that $\sigma _i ^{(\pm)}(\pm 1) = \pm 1$. In order to see the connection with the previous sections we set $\nu = 1/q^2$ and find that $\sigma^{(-)}_i(e^{-\lambda}) \simeq \check{R}_{i,i+1}(\lambda,\eta)$ of \Ref{chRXXZasymp} and $\sigma^{(+)}_i(e^{\lambda/2}) \simeq \check{R}_{i,i+1}(\lambda,\eta)$ of \Ref{chRA22asymp}.

The irreducible representations of the BMW algebra $W_N(q, \nu )$ are more complicated than the irreducible representations of the symmetric group $\mathfrak{S}_N$ or the Hecke algebra $\mathcal{H}_N(q)$, although they can be parameterized by the Young diagrams \cite{BW,Jones}. The simplest, one-dimensional  irreducible representations of $W_N(q, \nu )$ are defined by the symmetrizer and antisymmetrizer, respectively. The symmetrizer of the $W_N(q, \nu )$ is given by
\begin{equation}
\label{Sym_N}
\mathcal{S}_N = \frac{1}{[N]_q!} \sigma _1^{(-)}(q^{-1}) \sigma _2^{(-)}(q^{-2}) \cdots \sigma _{N-1}^{(-)}(q^{-(N-1)})
\mathcal{S}_{N-1},
\end{equation} 
with $\mathcal{S}_1 =1$ and
\begin{equation}
\label{Sym_2}
\mathcal{S}_2 = \frac{1}{[2]_q} \sigma _1^{(-)}(q^{-1}). 
\end{equation} 
We use the standard notation for the q-factorial $[n]_q! = [n]_q  [n-1]_q\cdots [2]_q  [1]_q$ and  the q-numbers $[n]_q = (q^n - q^{-n})/(q-q^{-1})$. The elements $\mathcal{S}_n$, $n = 1, \dots , N$ are idempotents, i.e. $\mathcal{S}_n^2 = \mathcal{S}_n$ . In addition, the symmetrizer $\mathcal{S}_N$ is also central.

In the realisation on $\mathbb{C}^3 \otimes \mathbb{C}^3$ of the BMW algebra $W_2(q, q^{-2} )$
\begin{equation}
\label{sigmaproj}
\sigma_1 = \check{R} (\eta) = q P_5 - q^{-1} P_3 + \nu P_1, \quad  \nu = \frac{1}{q^2},
\end{equation} 
and $e_1$ is proportional to the rank one projector $P_1$
\begin{equation}
\label{e-P1}
e_1 = \mu P_1 = (q + 1 + q^{-1}) P_1.
\end{equation}
Thus  
\begin{align}
\sigma ^{(-)}_1(q^{-1}) &=  (q + q^{-1}) P_5, \\
\sigma ^{\pm 1}_1 P_5 &= q ^{\pm 1} P_5, \\
 e_1 P_5 &= 0.
\end{align}

Similarly, the antisymmetrizer of the $W_N(q, \nu )$ is given by 
\begin{equation}
\label{Antisym_n}
\mathcal{A}_N = \frac{1}{[N]_q!} \sigma _1^{(+)}(q) \sigma _2^{(+)}(q^2) \cdots \sigma _{N-1}^{(+)}(q^{N-1})
\mathcal{A}_{N-1},
\end{equation}
with $\mathcal{A}_1 =1$ and
\begin{equation}
\label{Antisym_2}
\mathcal{A}_2 = \frac{1}{[2]_q} \sigma _1^{(+)}(q) .
\end{equation}
The elements $\mathcal{A}_n$, $n = 1, \dots , N$ are idempotents and the antisymmetrizer $\mathcal{A}_N$ is also central in $W_N(q, \nu )$. Below we will show how to prove this statement, here we only notice that
\begin{equation}
\label{sigma(+)2}
\sigma _1^{(+)}(q) \sigma _1^{(+)}(q) = [2]_q \sigma _1^{(+)}(q).
\end{equation}

It is straightforward to see that
\begin{equation}
\label{Antisym_3}
\mathcal{A}_3 \simeq \sigma _1^{(+)}(q) \sigma _2^{(+)}(q^2)\sigma _1^{(+)}(q) 
= \sigma _2^{(+)}(q) \sigma _1^{(+)}(q^2)\sigma _2^{(+)}(q) .
\end{equation}

In the realisation (\ref{sigmaproj},13) 
\begin{align}
\sigma ^{(+)}_1(q) &=  [2]_q P_3, \\
\sigma ^{\pm 1}_1 P_3 &= - q ^{\mp 1} P_3, \\
 e_1 P_3 &= 0.
\end{align}
In addition, in this realisation (with $\sigma^{(+)}_i(e^{\lambda/2}) \simeq \check{R}_{i,i+1}(\lambda,\eta)$ of \Ref{chRA22asymp}), the antisymmetrizer $\mathcal{A}_3$ has rank one, as it was already noticed in \Ref{q-detA3}. Furthermore, a straightforward calculation yields $\mathcal{A}_4 = 0$. Consequently all the higher antisymmetrizers vanish identically, $\mathcal{A}_n \equiv 0$, for $n > 4$.

In a general case of $W_N(q, \nu )$, it can be shown that the following identities are valid 
\begin{align}
\label{sgma-Sn}
&\sigma _i^{(-)}(q)   \mathcal{S}_n =  \mathcal{S}_n \sigma _i^{(-)}(q) =  0 ,\\
\label{sgma+An}
&\sigma _i^{(+)}(q ^{-1})   \mathcal{A}_n =  \mathcal{A}_n \sigma _i^{(+)}(q ^{-1}) =  0 ,
\end{align}
for $i=1, \dots , n-1$ and $1< n \leqslant N$. The relations (\ref{sgma-Sn},25) can also be written in the following form
\begin{align}
\label{crsS-n}
&\sigma _i \mathcal{S}_n =  \mathcal{S}_n \sigma _i =  q \mathcal{S}_n ,\\
\label{creS-n}
& e_i  \mathcal{S}_n = \mathcal{S}_n e _i = 0 , \\
\label{crsA-n}
&\sigma _i \mathcal{A}_n =  \mathcal{A}_n \sigma _i =  -\frac{1}{q} \mathcal{A}_n , \\
\label{creA-n}
& e_i  \mathcal{A}_n = \mathcal{A}_n e _i = 0 , 
\end{align}
for $i=1, \dots , n-1$ and $1< n \leqslant N$. From these identities it is evident that $\mathcal{S}_N$ and $\mathcal{A}_N$ are central in $W_N(q, \nu )$. Also, using the relations (\ref{crsS-n}-29), it is straightforward to check that $\mathcal{S}_n$ and $\mathcal{A}_n$ are idempotents, i.e. $\mathcal{S}_n^2 = \mathcal{S}_n$ and $\mathcal{A}_n^2 = \mathcal{A}_n$, $n = 1, \dots , N$.

In the next section the BMW algebra $W_N(q, q^{-2} )$ will be used to describe the multiplet structure of the spectra of some open quantum spin chains.

%
%
\section{Open Spin Chain}

According to the quantum inverse scattering method the R-matrix $R(u,q)$ can be used to construct an auxiliary L-operator for an integrable spin system, identifying the two spaces of $R(u,q) \in \mathrm{End}(V\otimes V)$ as auxiliary and quantum space, respectively:
\begin{equation}
\label{L-mat}
L_{0j}(u) = R_{0j}(u,q).
\end{equation}
Notice that in this section we mainly use the multiplicative spectral parameter, which in the case of the model $XXZ_1$ is given by $u= \exp(- \lambda)$.
Then the monodromy matrix of a spin chain with N sites is the product of L-matrices in $\mathrm{End}(V_0)$ whose entries are in  $\mathrm{End}(V_j)$ \c{TakhFadI}
\begin{equation}
\label{T-mat}
T(u) = L_{0\, N}(u)L_{0\,N-1}(u)\cdots L_{0\,1}(u) ,
\end{equation}
while the entries of the monodromy matrix $T_{ab}(u)$ are operators on the whole space of states $\mathcal{H} = \otimes _{j=1}^N V_j$ (in the case under consideration $V_j=\mathbb{C}^3$).
As a consequence of the Yang-Baxter equation \eqref{YBE} for the R-matrix and \Ref{L-mat} one has \c{Faddeev,TakhFadII,KS1}
\begin{equation}
R_{00^{\prime}}\left(\frac{u}{w}\right) L_{0j}(u) L_{0^{\prime}j}(w) =  L_{0^{\prime}j}(w) L_{0j}(u)  R_{00^{\prime}}\left(\frac{u}{w}\right)
\end{equation}
and
\begin{equation}
\label{RTT}
R_{12}\left(\frac{u}{w}\right)T_1(u) T_2(w) =  T_2(w) T_1(u)  R_{12}\left(\frac{u}{w}\right),
\end{equation}
where $T_1(u) = T (u) \otimes \mathbb{1}$ and  $T_2(u) =  \mathbb{1}\otimes T (u)$ are operator valued matrices in the two auxiliary spaces $V_1 \otimes V_2$, written as elements of $\mathrm{End}(V_1\otimes V_2)$. 
The trace of the monodromy matrix $T(u)$ - the transfer matrix
\begin{equation}
\label{transfer-t}
t(u) = \mathrm{tr} _0 T(u) ,
\end{equation}
is the generating function of the integrals of motion, including the Hamiltonian, of the spin chain with the periodic boundary condition. 

In order to construct integrable spin chains with non-periodic boundary condition one has to use the Sklyanin formalism \cite{Sklyanin}. The corresponding monodromy matrix $\mathcal{T}(u)$ consists of the two matrices $T(u)$ \Ref{T-mat}
and a reflection matrix $K^{-}(u) \in \mathrm{End}(V)$
\begin{equation}
\label{open-T}
\mathcal{T}(u) = T(u) K^{-}(u) T^{-1}(u^{-1}).
\end{equation}
Using the unitarity relation \Ref{unRXXZ} ($R_{12}^{-1}(u^{-1})=R_{21}(u)$) one gets
\begin{equation}
\label{inv-T}
T^{-1}(u^{-1}) = R_{1\,0}(u)R_{2\,0}(u) \cdots R_{N\,0}(u) .
\end{equation}
Taking into account the definition $R_{12}(u,\eta) =\mathcal{P} _{12}  R_{21}(u,\eta)\mathcal{P} _{12}$ one can transform the monodromy matrix $\mathcal{T}(u)$ into the following form (in order to shorten the notation in the formulas below the argument $\eta$ will be dropped) 
\begin{equation}
\label{open-T(u)}
\mathcal{T}(u) = \check{R}_{N\,0}(u) \check{R}_{N-1\,N}(u) \cdots \check{R}_{1\,2}(u) K_1^{-}(u) \check{R}_{1\,2}(u) \check{R}_{2\,3}(u) \cdots \check{R}_{N\,0}(u) .
\end{equation}
The generating function $\tau(u)$ of the integrals of motion \cite{Sklyanin} is given by the trace of $\mathcal{T}(u)$ over the auxiliary space with an extra reflection matrix $K^{+}(u)$
\begin{equation}
\label{tau(u)}
\tau (u) =  \mathrm{tr} _0 \left( K_0^{+}(u) \mathcal{T}(u) \right).
\end{equation}
The reflection matrices $K^{\pm}(u)$ are solutions to the reflection equation with a property $K^{-}(1) = \mathbb{1} \in \mathrm{End}(V)$ and $\tau (1)\simeq \mathbb{1}$. In particular, the Hamiltonian is given by $H = \frac{1}{2} \frac{d}{du} \ln \tau(u) |_{u=1}$,
\begin{equation}
\label{Ham-open}
H =  \sum _ {i=1}^{N-1} \check{R}_{i,i+1}^{\prime}(1) + \frac {\mathrm{tr} _0 K_0^{+}(1)\check{R}_{N\,0}^{\prime}(1)}{\mathrm{tr} _0 K_0^{+}(1)} + \frac{1}{2} \left( \frac{dK_1^{-}(1) }{du} + \frac{1}{\mathrm{tr} _0  K_0^{+}(1)} \frac{d  \, \mathrm{tr} _0 K_0^{+}(1)}{du}\right) .
\end{equation}
The Hamiltonian density $h_{i,i+1} = \frac{d}{du}\check{R}_{i,i+1}(u) |_{u=1}$ as one can see from \Ref{RXXZ} is a function of the generators of $W_N(q, q^{-2})$  on the space $\mathcal{H} = \otimes _{1}^N \mathbb{C}^3$.
The two extra boundary terms are contributions from the two reflection matrices $K^{\pm}(u)$ at the sites $1$ and $N$.
In our case we can take the constant K-matrices $K^{-} (u)=1$ and $K^{+}(u)=Q^tQ$, where the matrix $Q$ is given by \Ref{Q-mat}. It is easy to check that a non-zero contribution at the site $N$ is proportional to the identity, hence it does not influence the structure of the spectrum. For general K-matrices the solution, by the algebraic Bethe ansatz, was given in \cite{KurakSantos}. 

Asymptotic expansion of $T (u)$ at $u \to 0$ (or at $u \to \infty$) results in some matrices which have no spectral parameter dependence in accordance with \Ref{RXXZ} (see also \Ref{chRXXZasymp}) 
\begin{equation}
\label{Tat0}
T (u) = u ^{-N} L_{0\,N}^-L_{0,N-1}^-\cdots L_{0\, 1}^-+ \mathcal{O}(u ^{-N+1}).
\end{equation}
Here the constant L-matrices $L_{0j}^-$ are upper triangular matrices which coincide with the asymptotic limit $\lambda \to +\infty$ \eqref{bgR} of the R-matrices \Ref{RXXZ},  $L_{0 j}^-=R_{0 j}^- = \mathcal{P} _{0 j}\check{R} _{0 j}$. 
Hence, the Yang-Baxter equation \eqref{YBE} for the constant R-matrix can be written as follows
\begin{equation}
\label{RLL}
R_{i, i+1}^- L_{0, i+1}^- L_{0 i}^- =  L_{0 i}^- L_{0, i+1}^-R_{i, i+1}^- .
\end{equation}
It follows from the formula \Ref{RXXZ} (and  also \Ref{chRXXZasymp}) that $R_{i, i+1}^- = \mathcal{P} _{i, i+1}\check{R} _{i, i+1}$. So, multiplying the previous equation by the permutation operator $\mathcal{P}_{i,i+1}$ from the left one gets
\begin{equation}
\label{chRLL}
\left[ \check{R}_{i, i+1},  L_{0, i+1}^- L_{0i}^- \right] = 0.
\end{equation}
It is then obvious that $\rho_W (\sigma_i) = \check{R}_{i,i+1} \equiv  \check{R}_{i}$, $\rho_W (e_i) = \mu \left(P_1(\eta) \right)_ {i,i+1}$ as the\break representation $\rho_W$ of the generators of the BMW algebra $W_N(q, q^{-2} )$ in the space $\mathcal{H} = \otimes _{1}^N \mathbb{C} ^3$, commute with the generators $T_{ab}^{-}$ of the global (or diagonal) action of the quantum algebra $\mathcal{U}_q(o(3))$ on the space $\mathcal{H}$
\begin{equation}
\left[ \check{R}_{i,i+1},  T^- \right] = 0, \quad T ^- = L_{0\,N}^-L_{0,N-1}^-\cdots L_{0\,1}^- .
\end{equation}
This product of $L_{0j}^-$ can be represented as the image of a multiple co-product map $\Delta ^N : \mathcal{U}_q(o(3)) \to \left(\mathcal{U}_q(o(3))\right)^{\otimes N}$ \cite{Drinfeld} acting on a universal L-matrix $\mathcal{L}_{0}^-$ with entries in $\mathcal{U}_q(o(3))$ on the representation space $\mathcal{H}$
\begin{equation}
T ^- = (\mathrm{id}\otimes \rho_W ) (\mathrm{id}\otimes\Delta ^N) \mathcal{L}_{0}^-.
\end{equation}
Analogously, the asymptotic expansion of $T (u)$ at $u \to \infty$ yields the matrix  $T ^+ = L_{0\,N}^+L_{0\,N-1}^+\cdots L_{0\,1}^+$ (cf. \eqref{Tat0}). Similar arguments used to show that $T^{-}$ commutes with $\check{R}_{i\,i+1}$ lead to the conclusion that 
$T ^+$ commutes as well.  Notice that the generators of the global action of the quantum algebra $\mathcal{U}_q(o(3))$ are entries of $T^{\pm}$. Analogous arguments are valid in the quantum algebra $A^{(2)}_2$ case as well.

It is known that in the space $\mathcal{H}$ as a space of representation of $\mathcal{U}_q(o(3))$ and $W_N(q, q^{-2} )$ these algebras are mutual centralizers \cite{SaB}. According to the centralizer property this induces the decomposition of the representation space $\mathcal{H}$ into direct sum of irreducible representations of both algebras, being a generalisation of the Schur-Weyl duality. Similarly to the Hecke algebra case, studied previously in \cite{KMN}, one gets
\begin{equation}
\label{decom}
\mathcal{H} = \sum _{s=0}^N V_s \otimes U_s ,
\end{equation}
where $V_s$ is the $(2s+1)$-dimensional irreducible representation of $\mathcal{U}_q(o(3))$ while $U_s$ is some irreducible representation of $W_N(q, q^{-2} )$. The dimension of an irreducible representation of $W_N(q, q^{-2} )$ is equal to the multiplicity $m$ of the corresponding irreducible representation of centralizer algebra $\mathcal{U}_q(o(3))$, and vice versa 
\begin{equation}
m(V_s) = \dim U_s , \quad m(U_s) = \dim V_s . 
\end{equation}
The dimension of the irreducible representation $V_s$ of $\mathcal{U}_q(o(3))$ and the number $n$ of the inequivalent irreducible representations in the decomposition \eqref{decom} %
are well known. It follows from the decomposition of the tensor product of the spin 1 representations of $o(3)$: $\dim V_s = 2s+1$,
\begin{equation}
n_N = N + 1, \quad m_N(V_s) = \sum _{j=s,s\pm1} m_{N-1}(V_j) , \quad s\neq 0, N-1,N,  
\end{equation}
together with $m_N(V_0) = m_{N-1}(V_1)$, $m_N(V_{N-1}) = 1 + m_{N-1}(V_{N-2}) = N-1$ and $m_N(V_{N}) = 1$. 
However,  the number and the dimensions of representations $U_s$ of $W_N(q, q^{-2} )$ can be obtained from its Bratteli diagram \c{BW, SaB}. For $N=2,3$ the number of existing irreducible representations of $W_N(q, q^{-2} )$ and those entering into the decomposition  of the space of states are the same $3,4$, respectively, while for $N \geqslant 4$ there are more irreducible representations of $W_N$ than of $\mathcal{U}_q(o(3))$, for example $n_4(W) = 8$ while $n_4({\cal U}_q(o(3))) = 5$.

The decomposition \eqref{decom} permits to determine the structure of the multiplets of the Hamiltonian, which is an element of the BMW algebra \Ref{Ham-open} being a function of the generators of $W_N(q, q^{-2} )$
\begin{equation}
\label{Ham-N-chain}
H = \sum_{i=1}^{N-1} h _{i,i+1} , \quad h _{i,i+1} = \frac{d}{d\lambda} \check{R} (\lambda, \eta) |_{\lambda =0} = f (\check{R}_{i}) \in W_N(q, q^{-2} ) .
\end{equation}


According to the QISM, the R-matrices \Ref{RXXZ} and \Ref{RA22} being regular at $\lambda = 0$ \Ref{regRXXZ} and \Ref{regRA22}, respectively, define the local Hamiltonian density for two sites of the corresponding spin chains \c{TakhFadI,KS}. For the $XXZ_1$-model from \Ref{chRXXZasymp} one gets
\begin{align}
\label{hamdensXXZ}
h_{XXZ} &= \frac{d}{d\lambda} \check{R} (\lambda, \eta) |_{\lambda =0} \simeq q \check{R} (\eta) - \check{R} ^{-1}(\eta)   \notag \\
&  = (q - 1) \left( (q+1+\frac{1}{q}) (P_5 - P_1) + P_3 \right).
\end{align}
In order to simplify the expressions of the eigenvalues the factor $(q - 1)$ will be dropped from the Hamiltonian density. 
In the $A_2^{(2)}$-case from \Ref{chRA22asymp} it follows
\begin{equation}
\label{hamdensA22}
h_{A} = \frac{d}{d\lambda} \check{R} (\lambda, \eta) |_{\lambda =0} \simeq q \check{R} (\eta) + \frac{1}{q^2} \check{R} ^{-1}(\eta) = (q^2 + \frac{1}{q^3}) P_5 + (1+\frac{1}{q}) (P_1 - P_3). 
\end{equation}
The Hamiltonian of the open spin chain with N-sites is then given by
\begin{equation}
\label{Ham-N}
H = \sum_{i=1}^{N-1} h _{i,i+1}. 
\end{equation}
As an example let us consider the case of $N=3$ sites when the algebra $W_3(q, 1/q^2 )$ is realised in $\mathbb{C}^3\otimes \mathbb{C}^3\otimes \mathbb{C}^3$ and the corresponding Hamiltonians are 
\begin{equation}
\label{Ham-3}
H =  h _{12} + h _{23}. 
\end{equation}
From the relations (\ref{hamdensXXZ}, \ref{sgma-Sn}, \ref{crsA-n}) it follows
\begin{align}
\label{}
    H_{XXZ} \mathcal{S}_3 &= 2(q+1+\frac{1}{q}) \mathcal{S}_3, \\
    H_{XXZ} \mathcal{A}_3 &= 2 \mathcal{A}_3
\end{align}
and similarly for the $H_A$ \Ref{hamdensA22}
\begin{align}
\label{}
    H_{A} \mathcal{S}_3 &= 2 (q^2 + \frac{1}{q^3})  \mathcal{S}_3, \\
    H_{A} \mathcal{A}_3 &= -2 (1+ \frac{1}{q})  \mathcal{A}_3 .  
\end{align}

In the case $N=3$ there are four irreducible representations (irreps) of $W_3$: two one-dimensional irreps generated by $\mathcal{S}_3$ and $\mathcal{A}_3$, respectively, the three-dimensional irrep $d_3$ (corresponding to the one-box Young diagram) and the two-dimensional irrep $d_2$ (corresponding to the three-box Young diagram with two rows). Thus the Hamiltonian being restricted to invariant subspaces can have up to seven distinct eigenvalues. Their multiplicities are obtained from the correspondence between the irreps of $W_3$ and the irreps 
of ${\cal U}_q(o(3))$: 
\begin{equation}
U(\mathcal{S}_3) \sim V_3, \quad U(\mathcal{A}_3) \sim V_0 \quad U(d_3) \sim V_1 \quad 
U(d_2) \sim V_2. 
\end{equation}
The degeneracies of corresponding energy values are
\begin{equation}
m(\epsilon(\mathcal{S}_3))=7,\  m(\epsilon(\mathcal{A}_3))=1,\ 
m(\epsilon_j(d_3))=3,\  m(\epsilon_k(d_2))=5,\ j=1,2,3;
\ k=1,2.
\end{equation}
The exact values of the corresponding energy are obtained by direct calculations and are given below. For the XXZ-model of spin 1 the corresponding expressions are
\begin{align}
\label{engXXZ}
    & \epsilon(\mathcal{S}_3)= 2 (q+1+\frac{1}{q}), \quad  \epsilon(\mathcal{A}_3) = 2 , \\
    & \epsilon_1(d_3)= 1,\quad \epsilon_{2,3}(d_3)=\left(\frac{1}{2}\pm\sqrt{\frac{1}{2}+ 2 (q+3+\frac{1}{q})}\right),\\
    & \epsilon_1(d_2) = (q+1+\frac{1}{q}), \quad \epsilon_2(d_2) = (q+3+\frac{1}{q}).
\end{align} 
In the $A_2^{(2)}$-case the corresponding expressions are
\begin{align}
\label{engA22}
    & \epsilon(\mathcal{S}_3)= 2 (q^2 + \frac{1}{q^3}), \quad  \epsilon(\mathcal{A}_3) = -2 (1+ \frac{1}{q}), \\
    & \epsilon_1(d_3)=  (q^2 + \frac{1}{q^3}), \notag \\ 
    & \epsilon_{2,3}(d_3)= \frac{1}{2}\left((q^2 + \frac{1}{q^3})\pm
       \sqrt{q ^4+ 8q ^2- 8q +\frac{34}{q}-\frac{8}{q^3}+\frac{8}{q^4}+\frac{1}{q^6}}\right),  \\
    & \epsilon_1(d_2) = (1+ \frac{1}{q})(q^2 -1+\frac{1}{q^2}), \quad \epsilon_2(d_2) = (1+ \frac{1}{q})(q^2 -2q + 1 - \frac{2}{q}+\frac{1}{q^2}).
\end{align} 
Although the Hamiltonian density \Ref{hamdensA22} has a common factor $(1+1/q)$ it is not convenient to drop it since the expressions of some eigenvalues become more cumbersome.

%
%
\section{Conclusions}
Two integrable spin systems invariant with respect to the quantum algebra $\mathcal{U}_q(o(3))$ were considered. 
These spin systems are defined in the framework of the QISM by trigonometric R-matrices related  to the 
quantum affine algebras $\mathcal{U}_q(\widehat{o(3)})$ and $\mathcal{U}_q(A_2^{(2)})$. It was shown that the mutually commuting integrals of motion belong to the image of the BMW algerba $W_N (q, q^{-2})$ in a reducible representation on the space of states ${\cal H} = \otimes_1^N \mathbb{C}^3$. The symmetry algebra and the BMW algebra centralize each other in the representation space, and this determines the structure of the spin system spectra. 

We point out that there is a series of quantum super-algebras $\mathcal{U}_q(osp(1|2n))$ \c{SaB} with corresponding R-matrices in the vector representation defining generators of the Birman-Wenzl-Murakami algebra $W_N(-q, -q^{-2n})$ similarly to the one considered in our paper. From the representation of the BMW algebra given by the R-matrix one can get two spectral parameter dependent R-matrices by the Yang-Baxterization procedure \Ref{sigma(u)}. Each of them yields solutions of the standard and ($\mathbb{Z}_2$ graded) super-Yang-Baxter equations \cite{KS1}. This results in a possibility to construct four series of integrable spin chains whose structure of the spectrum is similar to the one considered in the previous section.


%
%
\section{Acknowledgments}
We acknowledge useful discussions with V. ~Tarasov. 
This work was supported by RFBR grant 07-02-92166-NZNI\_a, 08-01-00638 and the FCT project \hfil\break
PTDC/MAT/69635 /2006.

%
%


\end{document}